# Recent Advances on HEVC Inter-frame Coding: From Optimization to Implementation and Beyond

Yongfei Zhang[1*], *Member, IEEE,* Chao Zhang[1], Rui Fan[2],
Siwei Ma[3], *Member, IEEE*, Zhibo Chen[4], *Senior Member, IEEE, C.-C. Jay Kuo*[5], *Fellow, IEEE*

*Abstract*—High Efficiency Video Coding (HEVC) has doubled the video compression ratio with equivalent subjective quality as compared to its predecessor H.264/AVC. The significant coding efficiency improvement is attributed to many new techniques. Inter-frame coding is one of the most powerful yet complicated techniques therein and has posed high computational burden thus main obstacle in HEVC-based real-time applications. Recently, plenty of research has been done to optimize the inter-frame coding, either to reduce the complexity for real-time applications, or to further enhance the encoding efficiency. In this paper, we provide a comprehensive review of the state-of-the-art techniques for HEVC inter-frame coding from three aspects, namely fast inter coding solutions, implementation on different hardware platforms as well as advanced inter coding techniques. More specifically, different algorithms in each aspect are further subdivided into sub-categories and compared in terms of pros, cons, coding efficiency and coding complexity. To the best of our knowledge, this is the first such comprehensive review of the recent advances of the inter-frame coding for HEVC and hopefully it would help the improvement, implementation and applications of HEVC as well as the ongoing development of the next generation video coding standard.

## I   Introduction

HEVC, also known as H.265 or MPEG-H Part 2, is the latest video coding standard developed by JCT-VC (Joint Collaborative Team on Video Coding) [1]. Compared with H.264/AVC[2], HEVC provides significantly improved coding performance of as much as 50% bit rate reduction at equal perceptual quality, with the employment of many efficient coding tools, including quad-tree based coding unit (CU)/prediction unit (PU) structure, advanced motion vector prediction (AMVP), etc. [1]. On the other hand, the performance is achieved at up to 4-10 times higher complexity than H.264/AVC and up to 5000 times to real-time video applications [3, 4]. This poses great challenge and draws great attentions from both academic and industrial community.

Among all coding techniques, inter-frame coding is the most important and efficient one. It finds the best matched block in reference frames to reduce temporal redundancy, the major one in video compression, between successive frames. Then, only

Y. Zhang and C. Zhang are with the Beijing Key Laboratory of Digital Media, School of Computer Science and Engineering, and State Key Laboratory of Virtual Reality Technology and Systems, Beihang University, Beijing, China, 100191. R. Fan is with China Academy of Electronic and Information Technology, Beijing, China, 100041. S. Ma is with the Institute of Digital Media, School of Electronics Engineering and Computer Science, Peking University, Beijing 100871, China. Z. Chen is with University of Science and Technology of China, Hefei, Anhui, 230026, China. C.-C. Jay Kuo is with University of Southern California, Los Angeles, CA 90089, USA. This work was partially supported the National Natural Science Foundation of China (No. 61772054, 61632001).
Copyright © 2019 IEEE. Personal use of this material is permitted. However, permission to use this material for any other purposes must be obtained from the IEEE by sending an email to pubs-permissions@ieee.org.
*Corresponding author: Yongfei Zhang, E-mail: yfzhang@buaa.edu.cn.

Motion Vectors (MVs), generated by Motion Estimation (ME) and representing the displacement between the best matched block and the current block, and the residual after Motion Compensation (MC), instead of the original pixels, need to be encoded and stored/transmitted. Inter-frame coding helps greatly remove the most significant temporal redundancy among consecutive frames, which also contributes more than two thirds of the overall computational complexity of HEVC, making it quite difficult for real-time applications. Therefore, it is necessary to reduce the complexity thus speed up the inter-frame coding in HEVC.

The inter-frame coding in HEVC can be roughly divided into three parts, i.e., CU/PU partitioning, Motion Estimation and Motion Compensation. *First*, the current encoding frame is divided into CTUs(Coding Tree Units) of size 64×64, and CTUs are further partitioned into CUs/PUs in a quad-tree structure, which later serve as the basic unit for subsequent inter/intra coding, as will be elaborated in II.A. *Second*, ME is performed on the basis of PUs in three steps. MV prediction first predicts the start search position for the following ME and integer-pixel ME are then employed to find the best-matched block while sub-pixel ME is finally conducted to obtain the final best-matched sub-pixel position. Please refer to II. B and II.C for the details. *Third*, MC is conducted and the residual between the original PUs and the best matched PUs in the reference frame(s) are generated, which is used to be encoded into bitstreams and stored or transmitted.

Although there have been few brief surveys on HEVC inter-frame coding, inter CU selection [5] or motion estimation [6-9] more specifically, these studies focused only on certain specific topics of inter-frame coding techniques and discussed only limited papers of the literature, while many new related algorithms have appeared in more recent years.

In this paper, we present a comprehensive survey of the inter-frame coding in HEVC and beyond, including fast inter-frame coding solutions, implementation on different hardware platforms as well as advanced inter-frame coding techniques. To the best of our knowledge, this is the first such kind of survey that extensively reviews the latest research advances of inter-frame coding, with more than 200 reviewed literatures. Hopefully it may provide valuable leads for the improvement, implementation and applications of HEVC as well as ongoing development of the next generation video coding standard.

The remainder of the paper is organized as follows. Section II provides a brief overview of the inter-frame coding in HEVC. In Section III and IV, the fast inter coding solutions and the implementations of inter coding on different platforms are reviewed. Section V surveys the recently developed powerful new inter frame coding schemes and Section VI concludes the paper.

## II  BRIEF OVERVIEW OF INTER-FRAME CODING IN HEVC

Before diving into recent advances on inter-frame coding, this section provides a brief overview of the original inter-frame coding techniques in HEVC video coding standard [1].

### A. CU/PU Partitioning

CTU is employed in HEVC, which can be much larger than traditional 16×16 MacroBlocks(MBs) in H.264/AVC [1, 2]. One CTU starts from 64×64 sized Largest Coding Unit (LCU) and recursively searches 4 levels of quad-tree depth up to 8×8 sized CUs. First, for a 64×64 sized CU, a prediction mode is determined with maximum compression efficiency. The current best CU of size 64×64 is compared to the CUs of the lower depth as shown in Fig. 1(a). That is, a function that finds the prediction mode for 32×32, 16×16, and 8×8 CUs is called recursively, and then the optimized CU depth is determined by comparing the cost of current best partitioning with that of the lower depths. The prediction mode is divided into the INTRA and INTER modes. INTRA mode uses the spatial locality for data compression, and the optimal PU mode is searched only using 2N×2N and N×N as in Fig. 1(b). It finds luma/chroma angles within certain search range that is most similar to the current PU. The INTER mode utilizes temporal locality for data compression, and finds the optimal prediction mode among 8 modes as in Fig. 1(b). In each PU mode, MV is derived through ME and used for encoding, which will be elaborated in subsequent subsections.

In all, for each CTU, in order to find the optimal CU/PU partition, Rate-Distortion Cost (*RDcost*) is recursively computed from 64×64 LCU to 8×8 CUs. And the best CU/TU partitioning is obtained through RDO (rate-distortion optimization)-based exhaustively search for all partitioning and prediction modes. As a result, a coding efficiency is much improved compared to the previous standard, but the computation overhead is also highly increased [3, 4]. Therefore, plenty of fast algorithms have been proposed, as will be reviewed in Section III. A.

B.  *Motion Estimation and Compensation*

ME/MC, the major contributor to video compression efficiency, finds the best matched block in reference frames to reduce temporal redundancy between successive frames. MV, representing the displacement between the best matched block and the current block, is generated by ME.

The entire ME process is made up of three coarse-to-fine procedures, namely, MV prediction, integer-pixel ME (IME) and Fractional-pixel ME (FME). *First*, MV prediction predicts the start search position for the following motion search using neighboring motion information. In HEVC, AMVP is adopted, which derives several most probable candidates based on data from adjacent PBs and the reference picture(s). The displacement between the starting search position and the current coding PU is called a predictive motion vector (PMV). HEVC also introduces a merge mode to derive motion information from spatially/temporally neighboring blocks[1].

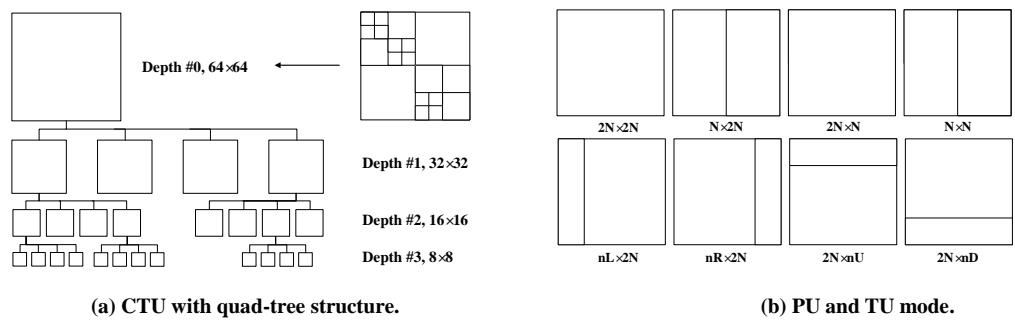

(a) CTU with quad-tree structure.   (b) PU and TU mode.

**Fig. 1. Quadtree structure-based CTU/CU/PU Partitioning in HEVC.**

The second step is IME, which conducts appropriate search strategies from PMV(s) until the best integer-pixel search position is obtained. Block matching algorithm (BMA) is the most popular algorithm for ME due to its simplicity, which determines the most matched block within a search window in reference frame(s) based RDO. The most straightforward full search (FS) traverses all positions in search window and obtains the optimal MV with the minimum *RDCost*. Although FS provides the best quality, its complexity is very high, as much as 40-80% of the total encoding time. To address this drawback, test zone search (TZS) is implemented as the build-in fast search mechanism in HM [10]. First, the start search position is determined by checking PMV and zero motion. Second, diamond search/square search is implemented, and an additional raster search is performed when the difference between obtained ME and start position is too large. Finally, an extra diamond/square search is performed as a refinement until the best search position is picked. To further reduce the complexity of TZS, plenty of fast IME algorithms have been developed, which will be reviewed in Section III. B.

*Third*, to further reduce the prediction residual, FME is implemented around the optimal IME position to obtain the final best sub-pixel position. In HEVC, 1/4-pel precision is used for MVs in luma component, and 7-/8-tap filters are used for interpolation of fractional sample positions (compared to 6-tap for 1/2-pel and linear interpolation for 1/4-pel in H.264/AVC). The interpolation for chroma components is similar to luma component, except that the number of filter taps is 4 and the fractional accuracy is 1/8 for the usual 4:2:0 format case. Thanks to the well-designed interpolation filters, HEVC improves the inter-coding performance by more than 10% over H.264/AVC[1]. On the other hand, FME accounts for approximately 60%~80% of the complexity of the entire ME process due to numerous interpolations and rate-distortion calculations[11]. It was also shown in [12] that the average number of pixel accesses, multiplications and additions during the interpolation in HEVC is almost twice of that in H.264/AVC. Consequently, plenty of fast FME algorithms have been proposed, which will be reviewed in Section III. C.

III  FAST INTER CODING SOLUTIONS

As overviewed in Section II, CU partitioning, integer- and fractional-pixel ME are the three most time-consuming modules in HEVC inter-frame coding. In this section, different fast algorithms are comprehensively reviewed and compared in terms of pros, cons, coding efficiency and complexity.

A. *Fast CU Partitioning*

To overcome the computational overhead for CU partition or depth decision, various fast algorithms have been proposed, which can be divided into three categories, namely top-down methods, bottom-up methods and prediction-based methods.

   (1)  **Top-down Methods**

HEVC defaults to a top-down visiting order to split a CTU to obtain the optimal CU depth decision. The top-down methods are most commonly used to divide a CTU with a tree-pruning approach to skip the CU partitioning process when the current CU partition is good enough in reducing redundant computing complexity. Early as the standardization of HEVC, several top-down methods have been introduced. Choi proposed a CU early termination algorithm that if the current optimal CU mode is skip, it will not to be recursively divided when a coding performance requirements have been met[13]. Gweon proposed to skip the PU search for the current CU depth if all Coded Block Flags (CBFs) of luma and chroma are zero, which had been adopted by HM [14]. Yang detected the SKIP mode early using differential motion vector and CBF[15]. In [16-21], if *RDCost* or the residual of a CU is smaller than a threshold, the recursive sub-CU will be stopped. However, since *RDCost* distributions of CUs are usually overlapped between the splitted and unsplitted CUs, it is often difficult to derive a precise threshold. [22-25] treat the CU partitioning as a probabilistic problem and determines whether the current CU need to be partitioned by comparing the probabilities of partitioning and non-partitioning. Bayesian-based CU size decision algorithms were proposed in [22, 23], which calculate the posterior probability of CU partitioning using Bayesian rule according to selected computational-friendly features. [24, 25] use KNN and Markov Random Field (MRF) respectively to derive the best CU partition.

(2) **Bottom-up Methods**

Motivated by the observation that CU sizes in complex scenes are usually close to the smallest CU while the top-down methods take a lot of computation in visiting large CUs, bottom-up methods are proposed with a reverse visiting order. Blasi proposed to accelerate CU partitioning with a reverse visiting order, in which the reference information is extracted by visiting smallest CUs to make bottom-to-up decisions[26]. However, the acceleration in less-texture/motion region is poor. In order to improve this framework, Zupancic [27] proposed an enhanced depth prediction and reverse CU selection algorithm through a prediction stage which determines what visiting order should be used based on MV variance distance. The bottom-up methods intrinsically favor motion and texture-rich video regions and are less efficient for less-texture and less-motion scenes. Luckily, nice performance can be achieved through predictive or adaptive CU visiting order, as below.

(3) **Prediction-based Methods**

Different from both top-down and bottom-up methods reviewed above, which aim to reduce redundant computation complexity by pruning or early skip unnecessary CU depth, prediction-based methods focus on determining directly a precise CU depth or depth range based on depth information of spatial-temporally neighboring and/or parent CUs. Shen derived the CU depth range according to the spatial correlation between neighbor CUs and specific depth used in the previous frame, and early termination is used to skip ME on unnecessary CU sizes[28]. Liu used the standard deviation feature of spatio-temporal depths and edge gradient of current CTU to predict the best CU depth [29]. [30, 31] classified the current CTU as either simple, medium or complex CTU using depth information of the left, top and co-located CTUs and excluded CUs of 64×64 (8×8) for the complex (simple) CTUs. Although better coding quality is achieved, the complexity reduction is limited since only *RDCost* calculation of one CU layer is excluded. Besides, the alternative CU depth range is fixed which lacks flexibility for videos of diverse content [28-31]. [32, 33] calculated the maximum CU depth range for different CTUs adaptively, which, however, costs redundant computation in lower CU depth as the top-down methods. Zhang *et al.* [34, 35] classified CTUs into multiples similarity classes and multiple unnecessary depths might be exclude adaptively. Later, [36] estimated the candidate CU depth range by exploiting the correlation of CU depth among current CU and temporally co-located CUs and utilizing the maximum depth of the co-located CUs and CBFs of the current CU to reduce the accumulated errors.

In conclusion, the prediction-based methods overcome the shortcomings of top-down and bottom-up methods that the coding performance will be affected by dynamic changes of video content, yielding less coding efficiency degradation. Besides, CU depth, edge gradient and other spatio-temporal information can be employed to achieve precise prediction of CU depth thus more complexity reduction under negligible coding efficiency degradation. Performance of fast CU partitioning algorithms are listed in TABLE I for comparison.

B. *Fast Integer-pixel Motion Estimation*

Fast IME has drawn great attentions due to at least the following two reasons. First, it is very time-consuming and takes 40-60% of the total encoding time. Second, its accuracy has a large influence on the performance of subsequent FME [12]. The research on fast IME can be divided into three categories, namely, search pattern design, search window decision and early termination strategies.

(1)Search Pattern Design Algorithms

Different search patterns has been designed in many fast block-matching algorithms, like three step search, four step search, diamond search, hexagon search, etc. These patterns can greatly reduce IME complexity, which, however, might fall into local optimum and lead to severe coding loss. [37, 38] presented pentagon and rotating pentagon pattern respectively. Yang proposed a directional search with a square pattern [39]. A quadratic pattern was introduced in [40], using SAD (Sum of Absolute Difference) distribution to predict the start point of each step by a coarse-fine order. In above algorithms, the same search pattern is used for all PUs regardless of their motion characteristics. To this end, a motion classification-based adaptive search pattern algorithm was proposed [41]. By exploring the motion consistency and *RDCost* of neighboring blocks, PUs are classified into either motion-smooth, -medium or -complex PUs. Then different search strategies are carefully designed for PUs in each class. Superior performance over state-of-the-art fast IME algorithms is achieved in terms of both coding performance and complexity reduction.

(2)Search Window Decision Algorithms

Algorithms in this class aim to speedup ME by reducing the search window size, i.e., the number of search points by dynamically adjusting the size of the search window.

TABLE I. PERFORMANCE OF FAST CU PARTITIONING ALGORITHMS

| Method | Properties | Algorithms/Ref. | Anchor | Configuration | BD-Rate (%) | BD-PSNR(dB) | Time Saving (%) |
|---|---|---|---|---|---|---|---|
| Top-down Methods | Adopted by standard* | Gweon 2011 [14] | HM 3.1 | RA | 0.7 | - | 36.3 |
| | | Choi 2011 [13] | HM 3.1 | RA | 0.4 | - | 39.6 |
| | | Yang 2011 [15] | HM 3.1 | RA | 0.2 | - | 32.1 |
| | Threshold approach | Choi 2012 [16] | HM 3.0 | RA/LD | -0.52/-0.44 (△Bit rate,%) | -0.05/-0.03 (Δ PSNR) | 41/37 |
| | | Yoo 2013 [17] | HM 5.0 | RA(α=1.0/α=1.5) | 0.08/0.33 (△Bit rate,%) | -0.02/-0.07 (Δ PSNR) | 37.3/49.56 |
| | | Shen 2014 [18] | HM 10.0 | RA/LD | 0.68/0.88 | - | 49.12/51.8 |
| | | Ahn 2015 [19] | HM 12.0 | RA/LDB | 1.4/1.0 | - | 49.6/42.7 |
| | | Lee 2015 [20] | HM 10.1 | RA/LD | 2.99/2.46 | - | 69/68 |
| | | Xiong 2015 [21] | HM 9.0 | RA/LD | 2.00/1.61 | - | 58.4/52 |
| | Probabilistic approach | Shen 2012 [22] | HM 4.0rc1 | Avg.(RA, LD) | 1.88 | - | 41.4 |
| | | Kim 2016 [23] | HM 15.0 | Avg.(AI, LD, LDP, RA) | 0.79 | - | 50.9 |
| | | Xiong 2014 [24] | HM 9.0 | LDP | 1.9021 | -0.0673 | 42.81 |
| | | Xiong 2014 [25] | HM 9.0 | LDP | 1.3194 | - | 52.58 |
| Bottom-up Methods | Reverse approach | Blasi 2015 [26] | HM 12.0 | RA | 1.95/0.70 | - | 59.1/20.6 |
| | Adaptive reverse approach | Zupancic 2016 [27] | HM 12.0 | RA(Encoder 1/2/3), LDB(Encoder 1/2/3) | 1.9/3.1/3.6, 1.5/2.8/3.3 | - | 57.5/67.0/68.8, 51.7/62.2/64.7 |
| Prediction-based Methods | Fixed depth range | Shen 2013 [28] | HM 2.0 | RA/LD | 1.49/1.15 | -0.049/-0.037 | 42/41 |
| | | Liu 2015 [29] | HM 16.0 | LDP | 1.6 | - | 40.5 |
| | | Zhou 2013 [30] | HM 8.0 | AI/RA/LD/LDP | 0.16/0.63/0.57/0.56 | -0.01/-0.02/-0.02/-0.02 | 22.6/20.3/20.8/22.0 |
| | | Zhao 2015 [31] | HM 11.0 | AI/LDP/RA | 2.510/2.381/1.922 (Δ Bit rate, kbit/s) | -0.092/-0.082/-0.078 (Δ PSNR, dB) | 54/67.6/68.4 |
| | Adaptive depth range | Zhang 2013 [34] | HM 8.0 | LDP | 0.16 | - | 25 |
| | | Fan 2014 [35] | HM11.0 | LDP | 0.65 | - | 35.72 |
| | | Liu 2016 [32] | HM 10.0 | RA/LDP | 0.9918/1.0517 | - | 59.76/56.71 |
| | | Chen 2016 [33] | HM 16.9 | RA | 1.6 | - | 47.5 |
| | | Li 2017 [36] | HM 2.0 | RA/LDB | 1.3/1.1 | - | 56.3/51.5 |

*For fair comparison, the results are listed according to that in [27].

As known, the search window for H.264/AVC is 16 and is extended to 64 in HEVC. HM exploits a dynamic search window to adjust the range of IME by calculating the temporal distance between reference and encoding frame. Ko found that horizontal/vertical MV differences (MVDs) roughly satisfy a Laplacian distribution and proposed an adaptive search window design based on the hitting probability of MVDs [42]. However, Dai [43] found the MVD distribution in [42] is more similar to Cauchy distribution through extensive experiments and improves accuracy of search range prediction accordingly. Shen divided the motion into three kinds, namely homogeneous-, normal- and complex-motion, and used MVD distribution of adjacent blocks to predict their search range respectively[44]. Liao further introduced the MVD of the parent CU and established a linear relationship between the size of the search window and the MVD of CTU [45].

**(3) Early Termination Strategies**

The early termination strategies achieve acceleration by terminating all or part of the IME process in advance. These algorithms can be further divided into two sub-classes.

TABLE II. Performance of Fast Integer-pixel Motion Estimation Algorithms

| Category | Algorithms/Ref. | Anchor | Configuration | Reported Performance | | | |
|---|---|---|---|---|---|---|---|
| | | | | BD-Rate (%) | BD-PSNR(dB) | ΔME Time (%) | Time Saving (%) |
| Search pattern design | Parmar2014 [37] | HM 14.0 | -- | 0.358 | -0.00399 (△PSNR-Y) | 22.295 | 14.506 |
| | Jeong 2015 [38] | HM 14.0 | -- | 0.2464 | -0.0078 | ≈18 | -- |
| | Yang 2014 [39] | HM 8.0 | -- | 0.146 | -0.006 | 60.80 | 11.16 |
| | Gao 2015 [40] | HM 14.0 | -- | 1.2 | -0.06 | -- | ≈56 |
| | Fan 2017 [41] | HM16.0 | LDP/RA | 0.17/1.12 | -- | 82.47[1] | 12.47/20.25 |
| Search window decision | Dai 2012 [43] | FS in HM 3.0 | LDP($C_{prob}$: 98.4/98.8/98/99.2) | 0.4/0.4/0.3/0.2 | -- | --- | ≈95 |
| | Liao 2015 [45] | HM 10.1 | -- | -- | -0.067 | 81.4 | -- |
| Early termination strategy | Nalluri2015 [46] | TZS in HM 16.0 | LDP/LDB/RA | 0.511/0.390/ 0.394 | -0.020/-0.015/ 0.055 | 55.14/42.29/ 34.32 | 10.83/33.98/ 24.60 |
| | Medhat 2016 [47] | FS in X265 | -- | 0.89(△Bit rate,%) | -0.0105 (△PSNR) | -- | 45.92 |
| | Pan 2016 [48] | TZS in HM 12.0 | LD/RA | 0.55/0.86 | -0.020/-0.034 | 20.12/18.52 | 15.04/12.29 |
| | Hu 2014 [49] | HM 12.1 | LD | 0.97(BD-Rate-Y, %) | -- | 69.76 | 13.05 |
| | Hu 2013 [50] | HM 7.0 | LD | -- | -0.0241 (△PSNR) | 73.49 | 16.71 |
| Overall | Li 2015 [51] | TZS in HM 10.0 | Threshold:10 | 0.5(BD-Rate-Y, %) | -- | -- | 49 |

[1] the ME time saving is measured as average number of search points (*ASP*) for one ME [12], as compared to that of TZS in HM.

The idea of the first class is that if the coding performance of current search location is acceptable, the subsequent search will be terminated. [46, 47, 52-54] proposed to compute the *RDCost* threshold using the encoding information of spatio-temporal adjacent blocks and determine whether the performance is acceptable with comparison to a threshold. Pan [48] found that if the MVP of the parent CU is equal to 0, the IME of the child PU can terminate early and directly use the MVP predicted by AMVP. But these methods are easy to reduce significantly the encoding performance because of the inaccurate thresholding.

The algorithms in the second class reduce the complexity by skipping motion search in the impossible locations. Typical algorithms include successive elimination algorithm (SEA) [55], multilevel SEA [56], global SEA [57] and confidence interval based algorithms [49]. The basic idea behind SEA and its variations[55-57] is that if the triangle inequality, that the difference of the norm of the encoding and reference block should be smaller than the optimal SAD achieved so far, is not satisfied, ME should be terminated. The confidence interval based algorithms [57] formulate IME as a statistical inference problem and estimate the confidence interval of the RD-cost for a given confidence level. The algorithms in [51, 58-61] achieve better tradeoff between coding performance and complexity by designing different ME strategies for regions of different content characteristics. In all, a good fast IME algorithm should be content-adaptive and the IME algorithms based on region classification are promising in achieving a better performance and complexity tradeoff. The fast IME algorithms are listed in TABLE II.

### C. Fast Fractional-pixel Motion Estimation

FME aims to find the minimum *RDCost* at the 49 sub-pixel locations around the optimal IME location to further reduce prediction residual and improve the coding performance. Fast FME algorithms can be roughly divided into two categories, namely interpolation-based and interpolation-free algorithms.

**(1) Interpolation-based algorithms**

The interpolation-based algorithms aims to speed up FME by skipping the sub-pixel interpolation and ME at sub-pixel locations where the *RDCost* may be poor.

HEVC adopts a coarse-to-fine FME which can be split into two steps. The first step searches the eight 1/2 pixels around the best IME position and finds the position with the minimum *RDCost*. The second step takes this position and 8 neighboring 1/4 pixels to find the minimum *RDCost* as the final optimal IME position. In [62, 63], different search patterns or strategies are designed to simplify the 1/2 and 1/4 pixel FME. [64, 65] predicted 1-3 candidate positions using estimated *RDCost* of surrounding integer pixels and interpolated the candidate position. [66] classifies PUs according to spatio-temporal correlation of IME and limits the search accuracy of FME of different PU classes. Since there are still large amount of interpolation and *RDCost* calculation or estimation, the complexity is still high.

**(2) Interpolation-free Algorithms**

Differently, interpolation-free algorithms aim to directly estimate the exact optimal sub-pixel position using *RDCost* distribution of adjacent integer pixels without time-consuming sub-pel interpolation and *RDCost* calculation.

Since it has been found that error surface of the integral-pixel *RDCost* within one integral-pixel range around the best IME position exhibits obvious singularity [67], the error surface can be used to reflect the *RDCost* distribution within the fractional-pixel range and to estimate the optimal FME position. Typical models include the 5-term model in *Eqn*. (1) [67], the 6-term model in *Eqn*. (2) [68] and the 9-term model in *Eqn*. (3) [69, 70].

$$f_5(x, y) = Ax^2 + By^2 + Cx + Dy + E \qquad (1)$$

$$f_6(x, y) = Ax^2 + Bxy + Cy^2 + Dx + Ey + F \qquad (2)$$

$$f_9(x, y) = Ax^2y^2 + Bx^2y + Cxy^2 + Dx^2 + Exy + Fy^2 + Gx + Hy + I \qquad (3)$$

where *A,B,...,I* are the parameters of error surface model, which can be obtained by curve fitting the *RDCost* of the four/eight integer pixel positions around the best IME.

The 9-terms models achieve the best prediction accuracy, which also lead to the highest computational complexity. Moreover, the prediction accuracy of above models might severely degrade when *RDCost* of one or more of the eight integer pixel are significantly different from that of the other ones. To solve this problem, Zhang proposed to approximate the minimum *RDCost* location on the error surface with the minimum location of the valley curve of the error surface [71]. Although [71]

TABLE III. PERFORMANCE OF FAST FRACTIONAL-PIXEL MOTION ESTIMATION ALGORITHMS

| Category | Algorithms/Refs | Anchor | Configuration | Reported Performance | | | |
|---|---|---|---|---|---|---|---|
| | | | | BD-Rate (%) | BD-PSNR(dB) | △ME Time(%) | Time Saving(%) |
| Interpolation-based Fast ME algorithms | Sotetsumoto2013 [62] | HM 9.0 | QP: 20,24,28 | 3.119 (△Bit rate, kbit/s)[1] | -0.018 (△PSNR-Y)[2] | 51.334[3] | -- |
| | Dai 2012 [64] | HM 3.0 | Lowdelay-loco | -- | -0.033 (△PSNR)[4] | -- | 54.065[5] |
| | Dai 2012 [65] | HM 3.0 | Lowdelay-loco | 17.51(△Bit rate, kbit/s)[6] | -0.00083 (△PSNR)[7] | -- | 45.52[8] |
| | Jia 2016 [66] | HM 14.0 | RA | 0.02 (△Bit rate,%) | -0.01 (△PSNR) | 40.86 | 24.56 |
| Interpolation-free algorithms | Zhang 2010 [71] | HM16.0[9] | LDP | 4.30 | -- | ~91 | -- |
| | Dai 2013 [72] | HM 6.0 | LDP | 4.00 | -- | - | -- |
| | | HM16.0[9] | LDP | 3.31 | -- | ~91 | -- |
| | Zuo 2015 [73] | HM 11.0 | LDP | 3.4 | -- | 91.1 | -- |
| | | HM16.0[9] | LDP | 2.95 | -- | ~91 | -- |
| | Fan 2017 [12] | HM16.0 | LDP | 2.42 | -- | ~90 | -- |

1. Averaged difference between the proposed bit rate and HM 9.0 bit rate in TABLE III of [62].
2. Averaged difference between the Proposed PSNR and HM 9.0 PSNR in TABLE III of [62].
3. Average value of reduction rate in TABLE III of [62].
4. Average value of △PSNR in TABLE I of [64].
5. Average value of reduced time in TABLE I of [64].
6. Difference between the proposed bit rate and hierarchical search bit rate in TABLE I of [65].
7. Difference between the proposed PSNR and hierarchical search PSNR in TABLE I of [65].
8. Hierarchical search total encoding minus Proposed total encoding time and divided by hierarchical search total encoding in TABLE I of [65], and then seek the average.
9. This result is from Ref. [12] for fair comparison.

achieves in most cases better performance than those directly fitting the whole error surface, it might suffer since there might be differences in the optimal IME positions derived from different IME algorithms and *RDCost* of the surrounding integer pixel positions might be not always monotonic. So in [72], Dai proposed to modify the anomaly of the intermediate result and Zuo [73] limited the minimum *RDCost* calculation in the range of [-1,1] in case the quadratic curve may not satisfy a convex or a linear function. Although [71-73] can significantly reduce the complexity of FME, there are still some problems. First, the assumption that the valley of the error surface vary along either the x- or y-axis, may be not always true. Second, the implicit assumption that the three minimum points are located along one straight line does not always hold. To overcome these problems, a multidirectional parabolic prediction-based algorithm were proposed in [12] to better accommodate different valley trends of the error surface and the valley curve is decomposed by passing the three minimum points with two projection parabolas to better fit the distribution of the valley curve.

In summary, since the time-consuming fractional-pixel interpolation and most *RDCost* calculations are skipped, these algorithms can achieve much more time saving as compared to those algorithms in the previous category, about 90% more specifically, as shown in TABLE III. However, the performance loss is relatively higher. How to accurately model the shape of error surface and improve the prediction accuracy might be one of the future focuses in fast FME.

IV IMPLEMENTATIONS ON DIFFERENT PLATFORMS

As reviewed above, plenty of fast algorithms have been proposed to reduce the high complexity of HEVC for real-time applications. However, their acceleration is quite limited, which reduces the complexity by at most dozens of times and still cannot satisfy the real-time demands. So great attention has been paid also to video encoder/decoder solutions on hardware platforms with powerful computing capacity and low power consumption. These can be categorized according to the platform they use, inlcuding CPU, GPU, Field Programmable Gate Array (FPGA) and Digital Signal Processor (DSP). The recent progresses in each category will be presented below.

*A. CPU-based Implementation*

As for the implementation on CPU, we mainly focus on the parallelization strategy, adopted in HEVC, namely Slice, Tile and WPP [1, 74], as listed below in TABLE IV.

**Slice**, as identical as that in H.264/AVC, is a data structure of either an entire or a region of a picture, which can be decoded independently from other slices of the same picture [1]. Ahn introduced a slice-level parallelization on HEVC encoder with SIMD implementation, which achieves about ten times speedup [75]. [76] focused on choosing proper slice size to achieve a better load balance among slices and use OpenMP to help complete slice-based parallelization programming.

**Tiles** [1] are introduced in HEVC to increase parallel processing capability. With tiles, a picture can be partitioned into rectangular regions/groups of CTBs by vertical/horizontal boundaries [77], which can be encoded independently. Misra [78] proposed a tile-based region of interest coding algorithm to achieve more bit rate savings. [79-82] tried to balance the encoding load among multi-cores through adaptive tile size partition based on content characteristic heuristics. However, both slice and tile would inevitably lead to coding efficiency degradation due to the dependency destruction among CTUs on slice/tile boundaries.

**WPP (Wavefront Parallel Processing)** [1] is a new tool in HEVC. Since no coding dependences are broken, WPP provides fine-granularity parallelism within a slice and often achieves better coding performance as compared to slices and tiles[1]. Radicke [83] implemented an WPP-based multi-threaded HM 11.0 encoder. Chi [74] improved WPP with a Overlapped WaveFront (OWF) parallelization scheme, which tackles the drawback of reduced parallel efficiency due to increased or reduced number of threads in WPP and achieves a frame rate of more than 100fps at 4K resolution on a standard multi-core CPU. Chen presented an Inter-Frame Wavefront (IFW) approach, using CTB-level dependency as a parallelism criterion to guide the dependence reduction in WPP [84]. A large-scale HEVC encoder were implemented with a wave front-based high parallel solution based on IFW [85] and a 3D-WPP algorithm was also proposed to improve the parallelism of IFW [86]. Zhang combined WPP with multi-slices/frames to enhance the parallelism and developed a real-time HEVC encoder on a multi-core platform [87]. Besides, hybrid parallelism were also implemented for HEVC decoder[88, 89].

Performance comparisons of different parallel tools has also been investigated. Ahn compared the performance of slice- and tile-level parallel tools, concluding that slice-level parallelism achieves more time savings and less coding loss, due to easier load balancing [90]. Maazouz [91] compared WPP and Tiles on an open-source HEVC encoder Kvazaar, which focuses on enabling high-level parallel processing, and concluded that WPP is the best parallelization technique of HEVC. [92] also compared the

TABLE IV. Performance of Parallelization Strategies

| parallelism | Algorithms/Refs | Anchor | Configuration | BD-Rate (%) | BD-PSNR(dB) | Speedup |
|---|---|---|---|---|---|---|
| Slice | Ahn2014 [75] | HM 9.0 | RA/LD | 2.93/2.24 | -0.12/-0.09 | 68.13/68.65(time saving,%) |
| Tile | Misra2013 [78] | HM 9.2 | AI/RA/LDB/LDP | -2.2/-2.2/-5.4/-5.5 | -- | -- |
|  | Storch2016 [81] | HM 16.0 | Common Test Conditions | 3.014[1] | -- | 12.5661 |
| WPP | Radicke2014[83] | HM 11 | RA/AI/LD | 0.75/0.125/0.825[2] | 0.026/0.004/0.028[3] | 5.065/4.77/5.123[4] |
|  | Chi 2013 [93] | -- | 2160P X86-LP-4threads/X86-HP-8threads/Tilera-8threads | -- | -- | 51.9/115.7/31.7 (frame per second) |
|  | Chen2014 [84] | -- | IBBBP | -- | -- | 16.11[5] |
|  | Chen2016 [85] | WHPTB | IBBBP | -- | -- | 55.355[6] |
|  | Zhang2014 [87] | WPP | CTU32/2slices/IPPP/IPP on FHM10.0 | -- | -- | 78/40/80/176(MPS[7] improvement, %) |
|  | Wen2016 [86] | IFW | IPPPPP/IBBBBP | -- | -- | 2.06/1.945[8] |

1. Average of BD-rate loss in TABLE 3 of [81].
2. Average of proposed BD-Rate-Y in TABLE 1 of[83].
3. Average of proposed BD-PSNR-Y in TABLE 1 of [83].
4. Average of proposed speedups in TABLE 2 of [83].
5. Average of proposed speedup in TABLE V of [84].
6. Average of proposed speedup in TABLE XI of [85].
7. Maximum Parallel Scalability [87].
8. Average of proposed speedup in TABLE 2 of [86].

performance of tile and WPP in decoding speedup, and finds that tiles are faster and more scalable than WPPs with high thread count. Regard less of parallel tools, content-adaptive load balancing is always a problem. Besides, since GPU provides higher parallelism, hybrid GPU/CPU solutions might be promising.

*B. GPU-based Implementation*

Due to great computing power and parallelism, GPUs have been widely used for acceleration of computation-intensive processing, like fast inter coding and encoder framework.

GPUs are used to calculate SADs in the GPU-based fast CU partitioning solution in [94]. A fast ME with adaptive search range decision was implemented on GPU, in which GPU performs ME in parallel with the decided search ranges from CPU [95]. Jiang [96] calculated SAD of IME by GPU and GPU memories are carefully designed and utilized to optimize the interpolation of FME. Similarly, a bi-predictive ME was implemented on GPU in [97]. What's more, Luo [98] proposed a GPU-based parallelized ME, in which hierarchical parallelization of ME process is achieved in CTU, PU and MV layers, respectively. A GPU-based parallel ME algorithm was also developed in [99], which improve GPU utilization by dividing a frame into two sub-frames for pipelined execution.

To further improve coding efficiency, efficient cooperation between CPU and GPU should be carefully considered. Kao [100] and Wang [101] shared similar ideas in using GPUs to identify the best AMV of each mode and CPUs to refine the AMV set for the best MV in ME. Wang proposed an efficient CPU/GPU parallel HEVC encoder framework with GPU focusing on ME, interpolation and border padding, achieving 113 times speedup as compared to CPU-based encoders [102]. Xiao introduced a fast HEVC encoding framework with multi-core CPUs and GPUs, in which GPUs are used to estimate MVs for each CU/PU partitions while CPUs are used to leverage the MC costs to speed up CU/PU partitions[103]. On the other side, hybrid GPU plus

CPU architectures for HEVC decoder were also studied[104, 105]. The performance of GPU-based implementations are listed in TABLE V. As can be seen, with the implementations on CPU/GPU, together with fast algorithms, HEVC encoding can be much accelerated. To further improvement, efficient cooperation and balance between CPU and GPU should be carefully considered.

*C. FPGA-based Implementation*

FPGA, characterized by reconfiguration ability, cheap cost and quick deployment, make it good candidate for hardware implementation of HEVC, as shown in TABLE VI.

In [106-108], some works has been done on FPGA-based fractional-pixel interpolator optimization. [106] synthesized a multiplier-less high throughput architecture on Altera FPGAs, achieving real-time interpolation of 3840×2460 QFHD frames. [107] developed a low-energy fractional-pixel interpolation hardware for all PU sizes on a Xilinx Virtex-6 FPGA, which achieves 48% energy saving and a frame rate of 30fps for 2K resolution videos. Ghani improved the frame rate to 45fps for 2K resolution videos, with a High-level synthesis (HLS) tool of HEVC fractional-pixel interpolation algorithm [108].

Since SAD is one of the most used distortion metric in codec of which the total number of calculation with 8 PU sizes in a CTU would be as much as 1361[109], research have been done on FPGA-based SAD optimization[109-111]. FPGA-based SAD architectures are developed in [109] to trade off total delay and hardware resources. [110] presented a lowcomplexity FPGA architectures to compute SADs for all block sizes. [111] introduced an parallel SAD architecture, which makes full use of the 64 processing units in parallel to calculate SAD and achieves 30fps of 2K resolution videos on Xilinx Virtex-7 XC7VX550T. On the other hand, FPGA-based HEVC decoders have also been studied [112-114].

In all, as a popular hardware and bridge between coding algorithms to final ASICs, FPGA will be an important option for HEVC implementation.

*D. DSP-based Implementation*

Compared to above hardware platforms, DSPs has become ideal for fast implementation of variety of computational-

**TABLE V. PERFORMANCE OF GPU-BASED IMPLEMENTATIONS**

| Category | Properties | Algorithms/Refs | Anchor | Configuration | Reported Performance | | |
|---|---|---|---|---|---|---|---|
| | | | | | BD-Rate (%) | BD-PSNR(dB) | Time Saving(%) |
| GPU-based fast inter coding algorithm | Fast CU partitioning | Lin2016 [94] | HM16.6 | LDP, Thr=0.75[1], GTX Titan Black GPU | 2.016 | -0.027 | 67.17 |
| | Fast ME | Kim2014 [95] | HM10.0 | LDP, GTX 780 with 3GB DRAM | 1.2 | -0.04 | 40.3 |
| | | Radicke2014[97] | HM13 | RA and RA 10bit | <0.8(△Bit rate) | <0.027(△PSNR, dB) | 54.16 |
| | | Luo2019 [98] | HM16.2 | LDP, Tesla K40C | 0.52 | -- | >12.7(ME speedup) |
| GPU-based encoder framework design | CPU+GPU | Wang2013 [102] | X265 | NVIDIA Tesla C2050 | -- | ≈0.7(△PSNR, dB) | ≈113(speedup) |
| | | Wang2014 [101] | X265 | AMD R9 290x,>=1080p | -0.079(△Bit rate) | -0.003(△PSNR, dB) | 32.77(speedup) |
| | | Kao2016 [100] | X265 | FS, Tesla K40/GTC 960 | 3.106/3.105(△Bit rate) | 0.001/0.002(△PSNR) | 17.62/9.74(speedup) |
| | | Xiao2015[103] | HM10.0 | IPPP, 2 Xeon E5-2687W, GTX 690. | 1.5 | -- | 28.8/52(speedup) |

1. A parameter in IV-B of [94].

TABLE VI. Performance of FPGA-based implementations

| Category | Algorithms | Configuration | Reported Performance | | |
|---|---|---|---|---|---|
| | | | #Slices | FIR filters Frequency (MHz) | Total power (mW) |
| FPGA-based interpolation | Kalali2014 [107] | Xilinx Virtex 6 | 1557 | 200 | -- |
| | Ghani2016 [108] | Xilinx Virtex 6 | 4426 | 168 | -- |
| FPGA-based SAD optimization | Purnachand 2013[110] | Xilinx Virtex 5, Non-parallel/ Parallel SAD | 8577/ 9182 | 174.673/ 171.947 | 91.3/ 136.18 |
| | Nalluri 2014 [109] | Xilinx Virtex 5, 2-stage Parallel architecture | 11738 | 165.57 | 320.86 |

extensive algorithms including video coding for its following features: 1) DSP employs Harvard structure in which program and data space are separated with a dedicated hardware multiplier; 2) DSP adopts multi-stage pipeline and multiple instructions can be simultaneously executed in one clock cycle; 3) It has a short instruction cycle, high operational precision, small power consumption and other characteristics. Compared to abundant DSP implementations of H.264/AVC[115-119] and HEVC decoder [120-123], research on HEVC encoders implementation is relatively rare. The high complexity of HEVC encoders, which is tens of times larger than that of HEVC decodes as well as H.264/AVC, making it very demanding however challenging. [124] proposed a DSP-based HEVC encoder, which focuses on simplifying mode decision process to decrease the complexity and uses SIMD and data-level parallelism to optimize SAD/SSE on single core DSP TMS320C6678. In our earlier research [125], we also used the powerful SIMD instructions to improve the parallel capacity of core time-consuming coding modules in HEVC. In [126], Sun introduced a Markov Chain model based data pre-fetching algorithm to speed up ME on multi-core DSPs through improving efficiency of data access. In our recent work [127], a multi-core DSP-based highly-paralleled HEVC encoder solution was implemented. The overall structure of HEVC encoder was re-designed to well support the encoding parallelism and low-delay low-memory multicore data transmission mechanism was designed to reduce data access latency Besides, SIMD-based optimizations are also integrated. Up to the best of our knowledge, this is the first of this kind to implement comprehensively the HEVC encoder on a DSP platform. However, a performance loss of 0.93dB is non-ignorable. TABLE VII presents the performance of typical DSP-based HEVC encoders. As can be seen, more work are needed to achieve real-time HEVC encoders with ignorable coding performance loss. The integration of different levels of encoding parallelism as well as full utilization of the resources on multi-core DSPs might be the direction to fulfill the real-time HEVC encoders on DSP platforms.

## V   Advanced Inter Coding Techniques

As reviewed above, plenty of work has been done on either algorithm optimizations or hardware implementations to facilitate the HEVC applications. Meanwhile, a series of new techniques are also proposed to further enhance the coding efficiency beyond HEVC [122], among which, QuadTree plus Binary Tree block partitioning, affine transform-based ME and adaptive precision ME are the most promising inter-frame coding tools. Besides, powerful Deep Learning has also been introduced to inter-frame coding, as reviewed in Section V. D.

TABLE VII. Performance of DSP-based HEVC encoders

| Category | Properties | Algorithms | Anchor | Configuration | BD-Rate (%) | BD-PSNR(dB) | Speedup |
|---|---|---|---|---|---|---|---|
| Single core DSP | SIMD(SAD/SSE) | Kibeya2016 [124] | HM12.0 | TMS320C6678(1-core) | 0.54 | -0.004 | 24.1% (time saving) |
| | SIMD | Zhang2017 [125] | HM10.0 | TMS320C6678(1-core) | -- | -- | 1.59-6.56 |
| Multi-core DSPs | Data pre-fetching for ME | Sun2016 [126] | HM10.0 | TMS320C6678(8-core) | -- | -- | 1.07~1.23 |
| | CTU-level parallelism, data transmission optimization, SIMD | Jiang2018 [127] | HM10.0 | TMS320C6678(8-core) | -- | 0.93 | 6.79/465.50[1] |

[1]The speedup of 6.79 is computed against the optimized HM10.0 while 465.50 is against the original HM10.0.

A. *Quadtree plus Binary Tree (QTBT) Block Partitioning*

Block partitioning structure is regards as core infrastructure based on which sophisticated coding tools are supported [127]. HEVC quadtree-based block structure is recognized as one of the representative changes that largely outperform that in H.264/AVC [127]. However, there are still some limitations, based on which potential improvements can be attained. 1) CU can only be square following the quadtree structure; 2) PU has only a few fixed types, which restricts the potentials of the prediction; 3) the residuals can only be transformed in square shapes, which limits the potential of transform; 4) same block partitioning is applied to both luma/chroma components, the properties of which might, however, not be identical.

Facing above issues, QTBT block partitioning structure was proposed during recent VVC reference software JVET(Joint Video Experts Team) development [128, 129] and adopted in JVET[130] for its high efficiency. In particular, besides the quadtree structure in HEVC, QTBT provides recursive binary tree partition, which can be either symmetric horizontal or vertical splitting, to better adapt to the diverse video content.

QTBT is exampled in Fig. 2, where *MinQTSize*/ *MaxBTSize* are parameters to restrict the depths of QTBT. Moreover, QTBT structure of chroma/luma CTBs in I slices can be totally different while for P/B slices chroma/luma CTBs still share the same partitions. However, QTBT brought significantly increased complexity, 523% in All-Intra (AI) configuration [131], originating mainly from exponential growth of combinations of partition patterns.

In view of this, JVET-F0063 [132] skipped the partition process of the second child code block of a BinaryTree partition, when *RDCost* of the parent and its sub-blocks satisfy some heuristic constraints. In JVET-D0077[133] split decision and intra mode was re-used if the same block in the other partition choices has the same neighboring coded blocks. In [134], QTBT partition is decided by adaptive adjustment of *MaxBTDepth* for each frame based on temporal information. These methods highly depend on global statistics of previously coded CUs, which may be difficult to handle texture/motion heterogeneous CUs. A local-constrained QTBT scheme was proposed by dynamically deriving the partition parameters for each CTU [135]. A CNN-based fast QTBT algorithm was presented [136], in which QTBT depth range is modeled as a multi-class classification problem and the depth range of 32×32 blocks are predicted directly instead of being judged at each depth level. In [138], the partition of QTBT at CTU level were dynamically derived to adapt to the local characteristics without transmitting any overhead and at CU level, a joint-classifier decision tree structure was designed to eliminate unnecessary iterations and control the risk of false prediction. [137] establishes a motion divergence field-based RD model to estimate *RDCost* of each partition mode and a confidence interval based early termination is proposed to remove unnecessary partition iterations. The performance of QTBT partitioning algorithms are tabulated in TABLE XIII. Since the block partitioning is readily a classification problem, machine learning especially DL techniques might be potential solutions.

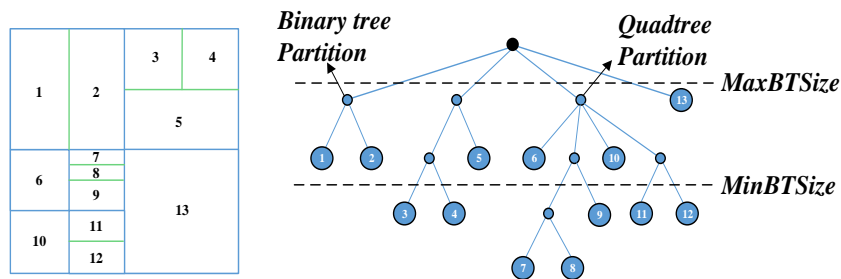

Fig. 2. QTBT block partitioning structure [137].

B. *Affine Transform-based Motion Estimation*

Block-based MEC serves as the fundamental technique to remove inter-frame redundancy[138]. The translational motion model has been used to characterize motions like panning and/or tilting for its simplicity and efficiency, which has, however, been demonstrated to be too simple to efficiently characterize complex motions as rotation, zooming and deformation [137, 139]. As early in 1993, Seferidis pointed out that high-order motion models, such as affine, bilinear, perspective models, are more efficient for complex motions than the translational one [140]. Among these high-order motion models, affine model has received the most attentiondue to its simplicity. As reviewed in [141], previous affine models can be divided into global and local models.

(1) Global Affine Motion Models

Global affine models were built to model motions between two frames with several groups of model parameters, and each reference frame is warped several times to generate multiple reference frames[141, 142]. Yu [143] used only one global model, and MVs of the salient features between the original and reference frame were used to determine the parameters for reference frame generation. For moving cameras-captured surveillance video coding, a global ME model is established by low rank singular value decomposition and adaptively clustering background MVs [144]. The global affine models are easy and straightforward, which, however, provides less accurate motion parameters for each local motion region, due to limited number of global affine motion models. Besides, the increased number of warped reference frames will significantly increase ME computation and bitrate overhead.

(2) Local Affine Motion Models

TABLE XIII. Performance Of Quadtree Plus Binary Tree (QTQT) Block Partitioning Algorithms

| Algorithms | Anchor | Configuration | Reported Performance | |
|---|---|---|---|---|
| | | | BD-Rate (%) | Time Saving(%) |
| Lin 2017 [132] | JEM 3.1[1] | AI Main | 0.06 | 12.1 |
| Huang2016 [133] | JEM 3.1[1] | AI Main | 0.05 | 11.44 |
| Yamamoto 2016 [134] | HM-13.0-QTBT[145][2] | RA | 0.52 | 18.7 |
| Wang 2016 [135] | HM-13.0-QTBT[145][2] | RA | 0.54 | 21.8 |
| | HM13.0 | RA/AI/LD/LDP | 0.6/0.5/0.7/0.6 | 35/28/26/22 |
| Jin 2017 [136] | JEM 3.1 | AI Main | 0.65 | 42.8 |
| Wang 2017 [138] | HM-13.0-QTBT[145] | RA/AI/LDB | 1.24/134/1.20 | 63.8/67.6/62.6 |
| | JEM 3.1[1] | AI Main | 2.67 | 31.8 |
| Wang 2018 [137] | HM-13.0-QTBT[145] | RA/LDB/LDP | 1.12/1.27/1.16 | 54.7/56.7/52.4 |
| | JEM-7.0 | RA/LDB/LDP | 1.23/1.37/1.27 | 50.6/53.1/50.2 |

Note: [1] This result is from [136] for fair comparison;
[2] This result is from [137] for fair comparison.

On the contrary, local models aims to build models for each local motion region, which can be further categorized into early-day mesh-based models and generalized block-based methods more commonly used nowadays.

In mesh-based method, a frame is divided into non-overlapping patches/meshes and the vertices are known as control points (CPs). The MVs of these CPs are used to determine the motions of other pixels through locally variant motion models (e.g., affine, perspective), with constraints to preserve the mesh structure. Then, the corresponding pixels in reference frame patches located by the generated motion information are warped into the current frame patch [146]. Toklu [147] added CPs hierarchically to better determine the motion models of each patch. [148] further proposed a content-based irregular mesh to better describe the object boundary. A merge mode for deformable block motion derivation was proposed in [146], in which the motion information of current block is derived from that of its neighboring blocks using bilinear interpolation model, six- or four-parameter affine model, with no ME and MV.

Mesh-based methods may provide a more truthful motion representation due to flexible shapes of meshes and the constraints. However, since the CPs are shared by neighboring blocks, it is hard to determine their MVs through ME in a block-based RDO process. Besides, due to various block sizes, 64×64 to 8×8 in HEVC and 128×128 to 8×8 in VVC(Versatile Video Coding, or H.266)[149], the problem of determining the MVs of CPs through RDO become even severe. Therefore, the mesh-based methods cannot be well integrated into the current video coding standards [141].

On the other hand, in generalized block-based methods [140, 150], each block can determine its own affine motion parameters, which is consistent with the standard video coding framework, except that the MVs are replaces by affine motion parameters for MC. Generalized block-based affine models are intuitively promising since it can better characterize complex motions, thereby improve coding efficiency. However, both ME and MC under affine motion models are significantly more complex conventional block matching-based ME/MC.

[151] aimed to derive a better prediction block through affine motion using surrounding translational MVs. Cheung [150] added an affine mode into the mode decision process and used the neighboring information to estimate the affine motion parameters of current blocks. [152] found that the affine motion model was more suitable for large blocks in HEVC. A fast gradient based affine ME scheme was proposed to decrease the encoder complexity [153]. Huang [154] extended the work in [150] for HEVC and designed the affine skip/direct mode to improve the coding efficiency. This was further developed to a quite complex affine MC framework [140], in which coding modes including affine inter/skip/ direct/merge were designed to fully exploit the motion correlation between neighboring blocks. Chen further developed the affine skip/direct mode to incorporate the merge mode for translational motion model and added temporal motion candidates into the candidate lists[155]. Affine motion compensation has been recently adopted in JVET [156]. Since affine schemes attempted to regenerate new affine models through motion of neighboring blocks which may correspond to different objects or have different motions, the regenerated affine motion model might be inaccurate.

In a few literatures, both global and local motion model were jointly considered. In [157-159], zoom motion was classified into two categories, i.e., global and local zoom motion. [159] introduced a affine and/or homographic projections-based global motion mode and a locally adaptive warped motion mode, to more accurately capture global motion at the frame level, or local motion at block level based on local motion statistics.

Although improved R-D performance might be achieved, the significant increase of ME complexity and more MVs hinders the adoption of affine model and other high-order models. A fast ME algorithm for zooming motion was presented in [160], which is, however, not applicable to more general cases. A simplified four-parameter affine model-based coding framework was proposed [141], which reduces the number of affine motion parameters from 6 to 4. Besides, an advanced affine MV prediction and an affine merge scheme were proposed to further encode the affine motion parameters.

Performance of affine transform-based ME algorithms are listed in TABLE IX. Since high-order global motion models are sometimes more effective to model the global motion for all local regions, while local models are more effective to precisely characterize the subtle motion of local regions, how to integrate the high-order global and local motion models into a whole framework might be a promising future work.

C. *Finer Precision Motion Estimation and Compensation*

MV, the relative displacement between the current and prediction block in reference frame(s) in MEC, explicitly describes the motion activity in the consecutive frames. MV resolution plays a key role in MEC and has significant impact on the coding efficiency [161, 162]. More specifically, increased MV resolution may lead to better prediction and smaller residual energy, thus

TABLE IX. Performance of Affine transform-based Motion Estimation Algorithms

| Algorithms | Anchor | Configuration | Reported Performance | |
|---|---|---|---|---|
| | | | BD-Rate (%) | Time (%) |
| Huang 2012 [154] | HM 1.0 | LL/RL[1] | -1.8/-1.7 | -- |
| Huang 2013 [140] | HM 1.0 | LL/RL/LH/RH[1] | -8.8/-5.0/-4.8/-2.1[2] -27.4/-19.1/-12.5/-4.1[3] | -- Encoding:154/168/152/165 |
| Narroschke 2013 [152] | HM 7.0 | LD | -2.4 | -- |
| Heithausen 2015 [139] | HM 12.1 | RA/LD | -0.9/-0.9[4]  -2.5/-2.9[5] | -- |
| Chen 2015 [155] | HM 12.0 | RA/LDB/LDP | -1.4/-.2.2/-3.5 | Encoding: ~120;   Decoding: ~130 |
| | Huang 2013 [139] | RA/LDB/LDP | -1.0/-1.0/-0.8 | -- |
| Li 2015 [153] | HM 11.0 | RA/LDB/LDP | -0.3/-0.5/-1.1 | Encoding: 419/323;   Decoding: 507/951 |
| Zhang 2016 [146] | HM14.0 | RA/LDB/LDP | -1.3/-1.7/-2.7[6] | Encoding: 116/111/112; Decoding: 11/114/115 |
| | HM14.0 | RA/LDB/LDP | -11.0/-15.2/-17.4[7] | Encoding: 119/113/113; Decoding: 121/137/13 |
| Li 2017 [141] | HM 16.7 | RA/LD | -1.0/-1.5[8] | Encoding: 118/128;   Decoding: 103/105 |
| | HM 16.7 | RA/LD | -11.1/-19.3[9] | Encoding: 121/131;   Decoding: 112/123 |
| Parker 2017 [159] | AV1[141] | -- | -3.0 | -- |

Note:  [1] LL/RL/LH/RH denote low-delay low complexity, random-access low complexity, low-delay high-efficiency, and random-access high-efficiency configurations in early HM test conditions.
[2] This is the averaged performance on common HM test sequences while [3] this is the performance on BQSquare sequence only.
[4]This is the performance with affine motion model while [5] this is the performance with zooming&rotating model.
[6] This is the averaged performance on common HM test sequences while [7] this is the performance on deformation sequences.
[8] This is the averaged performance on common HM test sequences while [9] this is the performance on affine test sequences.

improved coding efficiency. Integer-pel MV resolution was first used as early in H.261, and 1/2-pel resolution was introduced in MPEG-2 and H.263 while 1/4-pel resolution was adopted since H.264/AVC[2] and remained in HEVC [1]. During the development of HEVC, 1/8-pel MV resolution was also investigated [163], which was, however, not finally adopted due to increased complexity and limited coding gain. First, finer MV resolutions evoke increased complexity in sub-pel interpolations and MEC in both encoder/decoder. Second, more bits are required to represent the higher resolution MVs. Thus, it is fairly straightforward that a nice tradeoff should be found for MV resolution to balance the accuracy of MEC for higher coding gains, the bit-budget used for signaling MVs, and the encoding complexity [161, 164]. Thereafter, in HEVC, MV is still represented using 1/4-pel resolution and is uniform across all blocks/frames. However, the fixed and uniform MV resolution might not be optimal due to varying characteristics among different blocks. This inspires researchers for fine and adaptive MV resolution schemes, which can be roughly divided into two categories, namely block-based ones and pixel based ones.

**(1) Block-based Schemes**

A progressive MV resolution(PMVR) scheme was proposed in [165], where the MV resolution is progressively adjusted on the basis of MVD's magnitude. Later, PMVR was improved using adaptive thresholding based on CU depth [166], i.e, finer (coarser) MV resolution for small (large) CU depth. [167] improved PMVR by adaptation of MV resolution using PU size, gradient, MV components and spatiotemporal characteristics of the frames. [168] analyzed the potential influencing factors of optimal MV resolution, including texture complexity, motion scale, inter-frame noise and quantization parameter. An adaptive CU level MV resolution selection scheme was adopted in JVET [169], where the MV resolution can be either 1/4-pel or integer-pel, and a flag of the selected MV resolution is signaled to decoder for each CU. A adaptive MV resolution scheme was proposed based on a rate-distortion model[170], where an approximately linear relationship between the prediction distortion and MV resolution is observed and a rate model was built by dividing MVDs into three types according to their numerical values. A Cost-effective 1/8-pixel ME scheme was developed in [171], which jointly considered each fractional pixel's likelihood of being the optimal MV and compensation complexity. Later, [164] improved the PMVR [165] by using the size and the average gradient of a PU and a smarter ME algorithm around multiple MVPs is also introduced to further exploit the scheme. Finally in a very recent work, an adaptive PMVR scheme was proposed [161] to adaptively adjust the optimal MV resolution by decision trees constructed with a new RD model in terms of the MV resolution, significantly outperforming HEVC and PMVR with almost no computation overhead. As can be expected, insight investigation of the spatio-temporal video characteristics and relations could be employed to achieve a better tradeoff between accurate MV prediction and MEC complexity.

(2) Pixel-based Schemes

The above adaptive MV resolution schemes are all designed for block-based MEC, for its simplicity, robustness and ease of implementation. However, the drawbacks are also obvious. For instance, in case of tiny motion within a PU, the actual MV of each pixel might differ and it might fail to find an accurate match for blocks containing edges and textures. To this end, a more efficient and precise inter prediction approach is desired to accurately predict the MV of each pixel. [172] proposed a motion vector fields (MVF)-based coding scheme, which formulates MVF estimation as a discrete optimization problem of both the residual energy and MVF smoothness. However, this results in higher complexity, mainly coming from the extra MVF compression. Similarly, Alshin [173] proposed a pixel-wise motion refinement method called Bi-directional Optical flow (BIO), by combining optical flow concept and high accuracy gradients evaluation. As a combination of block- and pixel-wise MC, this method can handle tiny movements within a block effectively, and hence, remarkable coding gain can be achieved. In [174], a strict mathematical derivation to BIO was provided. [175] extended BIO to PUs with two unidirectional reference blocks, with a median filtering to MV shifting values of neighboring pixels to get the robust MV for each sample while a PU level pixel-wise motion refinement to further improve the coding performance.

The performance of latest finer and adaptive precision ME schemes are listed in TABLE X for reference and comparison.

D. Deep Learning-based Inter Coding

It is not new to introduced machine learning into video coding and a bunch of work has been done. Since it can be easily model as a classification problem, fast CU partitioning might be the first task in which machine learning are used, including Bayesian [22, 23], Markov Random Field [25], K Nearest Neighbors [24], Support Vector Machine[176-178] and Decision Tree [179, 180].

Recently, Deep Learning (DL) has emerged as a powerful machine learning tool and achieved great success and breakthroughs in the multimedia community, from high-level computer vision tasks, such as image classification, object detection, to low-level image de-noising and super-resolution, etc. More recently, DL has also been introduced to video coding

TABLE X. Performance of adaptive precision motion estimation Algorithms

| Category | Algorithms | Anchor | Configuration | BD-Rate (%) | Time (%) |
|---|---|---|---|---|---|
| block-based schemes | Ma 2013 [165] | HM 8.0 | RA/LDB/LDP | -1.7/-1.4/-3.8[1] | Encoding: 101/97/103;  Decoding: 100/101/101 |
| | Cho 2014 [166] | HM 7.0 | LDB/LDP | -1.6/-3.2 | Encoding: 92/97;         Decoding: 97/97 |
| | | PMVR [165] | LDB/LDP | -0.9/-0.9 | Encoding: 115/106;     Decoding: 97/97 |
| | Ray 2019 [167] | HM-16.6 | RA/LDB/LDP | -1.2/-2.7/-1.0 | Encoding: 132/135/131 Decoding: 118/119/119 |
| | | PMVR [165] | RA/LDB/LDP | -0.4/-0.6/-0.4 | Encoding: 125/125/122 Decoding: 116/118/118 |
| | Wang 2015 [168] | HM 16.2 | RA/LDB/LDP | -1.8/-1.7/-3.3 | Encoding: 93/92/101;    Decoding: 100/98/102 |
| | Wang 2016 [170] | HM 16.2 | RA/LDB/LDP | -1.5/-1.3/-2.5 | Encoding: 98/97/103;    Decoding: 96/94/101 |
| | Xiao 2017 [171] | full 1/4-pel search | -- | -1.3 | Encoding: 90.2 |
| | Ray 2017 [164] | HM 16.6 | RA/LDB/LDP | -1.2/-1.2/-3.2 | Encoding: 143/141/142; Decoding: 113/109/105 |
| | | PMVR [165] | RA/LDB/LDP | -0.5/-0.4/-0.5 | Encoding: 125/121/122; Decoding: 111/108/105 |
| | Wang 2017 [161] | HM 12.0 | RA/LDB/LDP | -1.3/-1.3/-2.4 -1.2/-1.2/-1.5[2] | Encoding: 97/98/102;     Decoding: 99/99/101 |
| | | PMVR [165] | RA/LDB/LDP | -0.6/-0.7/-0.7 | Encoding: 96/101/99;     Decoding: 99/98/100 |
| | | JEM 1.0 [169] | RA/LDB/LDP | -0.9/-0.8/-2.0 | Encoding: 103/103/105; Decoding: 102/102/101 |
| pixel-based schemes | Zheng 2015 [172] | HM 12.0 | LDP | -1.91 | -- |
| | Alshin 2010 [173] | [181] | Hierarchical B | -3.00 | -- |
| | Zhao 2016 [175] | HM 12.0 | LDB | -0.83 | -- |

Note: [1]This result is form [165] with parameters $TH_q$=4 and $TH_e$=2.
[2]The two sets of BD-Rate are under different training/testing conditions, while the encoding/decoding time is identical.

TABLE XI. Performance of DL-based Inter Coding Algorithms

| Algorithms | Anchor | Config | BD-Rate (%) | Time (%) |
|---|---|---|---|---|
| Yan 2017 [182] | HM 16.7 | LDP | -0.9 | -- |
| Xu 2018 [183] | HM 16.5 | LDP | 1.5 | 44-63 |
| Huo 2018 [184] | HM 12.0 | LDP | -2.3 | -- |
| Zhao 2018 [185] | HM 16.5 | RA | -3.1 | -- |
| Wang 2019 [186] | HM 16.9 | LDP | -1.7 | 3444 |
| Ibrahim2018 [187] | HM 16.9 | LDP | -2.6 | -- |
| Xia 2018 [188] | HM16.15 | LDP | -1.9 | -- |
| Zhao 2019 [189] | HM16.15 | RA/LDB | -3.0/-1.6 | 164.9/-- |

and dozens of work have been proposed for almost all modules along the coding/decoding process, including intra CU partitioning [190-193], block up-sampling for intra frame coding[194], intra mode decision [195-199], transform [200], rate control [201, 202], in-loop filtering/post-processing [203-205], arithmetic coding[206], or decoder-end artifact-removal and quality enhancement [207, 208].

However, only a few work has been done on inter coding. In [183], Convolutional neural network (CNN) and long-and short-term memory (LSTM) network are utilized to predict the CU partition for both intra-/inter-modes to reduce HEVC complexity. A CNN-based half-pel interpolation filter was proposed to address the problem of fixed interpolation filters for sub-pel MEC in HEVC[209]. A CNN-based MC refinement scheme was developed in [184], in which both the motion compensated prediction and the neighboring reconstructed region are used to train CNN models to help refine ME. In [186], a cascade neural network of a fully connected network (FCN) and a CNN is proposed for inter prediction algorithm in HEVC, in which the spatial neighboring pixels and the temporal neighboring pixels are fed to the network to perform accurate inter prediction. A fully connected networks-based interpolation-free FME scheme is proposed in [187], which utilizes the SSE of best IME and eight surrounding locations, PB height and width to predicts the best FME. In [188], a group variational transformation convolutional neural network (GVTCNN) were designed to improve the fractional interpolation performance of the luma component in motion compensation, which infers samples at different sub-pixel positions from the input integer-position sample. In [189], a CNN-based bi-prediction model was presented, in which the predictive signal can be automatically inferred in an end-to-end manner, and the non-linear mapping leads to better fusion in the bi-prediction process than conventional block-wise translational motion mapping. Chen proposed the concept of VoxelCNN [210], which includes motion extension and hybrid prediction networks, can model spatio-temporal coherence to effectively perform predictive coding inside the learning network. Please refer to two recent review papers [211, 212] for more deep learning-based image and video coding besides inter coding.

The performance of DL-based inter coding-related algorithms are listed in TABLE XI for reference. As can be seen, all these works have opened up a new direction that adopts DL into video coding to reduce the coding complexity or further enhance the coding efficiency. Besides, those coding modules with empirical or statistical assumptions, like interpolation filters, transforms cores for TUs, and in-loop filters, are all good candidates in which DL can be introduced for content-adaptive coding to achieve better coding efficiency. However, it should be also noted that the DL structures should be carefully designed for the coding tasks since real-time performance is essential for coding applications.

## VI Conclusions

In this paper, we presented a comprehensive review on the recent progress of inter-frame coding of the H.265/HEVC video coding standard. More specifically, the recent advances were classified into three categories, namely fast optimization solutions, implementation on different hardware platforms and advanced inter coding techniques. First, the fast solutions of the inter-frame coding techniques in HEVC, briefly introduced in Section II, were examined, which includes algorithms on fast CU partitioning, fast integral- and fractional-pixel ME. Then, implementations of HEVC inter frame coding on different platforms, icluding CPU, GPU, FPGA and DSP, were reviewed. Thirdly, we surveyed the recently developed powerful inter frame coding schemes, including QTBT block partitioning, affine transform-based ME, adaptive precision ME, as well as the emerging deep learning-based schemes.

Through such a comprehensive review of the recent advances of the inter frame coding in HEVC, hopefully it would provide valuable leads for the improvement, implementation and applications of HEVC, especially the ongoing development of the next generation video coding standard, at the current critical time of developing the next generation video coding standard beyond HEVC.

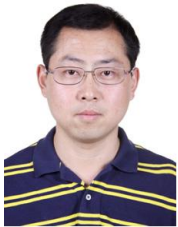

**Yongfei Zhang** (M'12) received the B.S. degree in Electrical Engineering and the Ph.D. degree in Pattern Recognition and Intelligent Systems, from Beihang University, Beijing, China, in 2005 and 2011 respectively. He is now an Associate Professor with the Beijing Key Laboratory of Digital Media, School of Computer Science and Engineering, Beihang University, Beijing, China. His current research interests include image/video analysis, compression, transmission, and video surveillance.

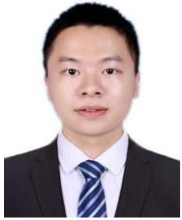

**Chao Zhang** received the B.S. degree in computer science from University of Electronic Science and Technology of China, and the M.S. degree in computer science from Beihang University, China, in 2019. After that, he joined Traffic Management Research Institute of the Ministry of Public Security, Wuxi, China. His current research interests include image/video processing and Internet of Vehicles.

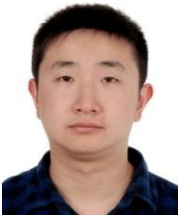

**Rui Fan** received the B.S. degree in Computer Science and Technology from Beijing University of Posts and Telecommunications in 2011. He received the Ph.D. degree with the Computer Science and Technology, Beihang University, Beijing, China, in 2017. His research interests are in the areas of video coding and its optimization on embedded systems. He is currently an engineer with China Academy of Electronic and Information Technology.

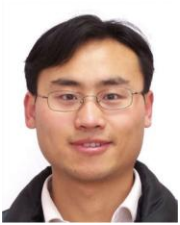

**Siwei Ma** (M'03–SM'12) received the B.S. degree from Shandong Normal University, Jinan, China, in 1999 and the Ph.D. degree in computer science from the Institute of Computing Technology, Chinese Academy of Sciences, Beijing, China, in 2005. From 2005 to 2007, he held a post-doctoral position at the University of Southern California, Los Angeles. Then, he joined the School of Electronic Engineering and Computer Science, Institute of Digital Media, Peking University, Beijing, where he is currently a Professor. He has published over 200 technical articles in refereed journals and proceedings in the areas of image and video coding, video processing, video streaming, and transmission.

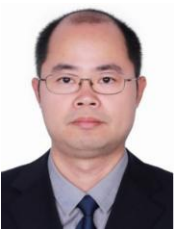

**Zhibo Chen** (M'01-SM'11) received the B.Sc. and Ph.D. degrees from the Department of Electrical Engineering, Tsinghua University, in 1998 and 2003, respectively. He was with SONY and Thomson from 2003 to 2012. He was a Principal Scientist and a Research Manager with the Thomson Research and Innovation Department. He is currently a Professor with the University of Science and Technology of China. His research interests include image and video compression, visual quality of experience assessment, immersive media computing, and intelligent media computing. He has more than 50 granted and over 100 filed EU and U.S. patent applications and more than 80 publications. He is a member of the IEEE Visual Signal Processing and Communications Committee and the IEEE Multimedia Communication Committee. He was an Organization Committee Member of ICIP 2017 and ICME 2013, and served as a TPC Member for IEEE ISCAS and IEEE VCIP.

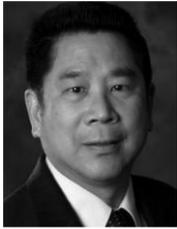

**C.-C. Jay Kuo** (F'99) received the B.S. degree in electrical engineering from National Taiwan University, Taipei, Taiwan, in 1980, and the M.S. and Ph.D. degrees in electrical engineering from the Massachusetts Institute of Technology, Cambridge MA, USA, in 1985 and 1987, respectively. He is currently the Director of the Multimedia Communications Laboratory and a Professor of electrical engineering, computer science and mathematics with the Ming-Hsieh Department of Electrical Engineering, University of Southern California, Los Angeles, CA, USA. He has coauthored about 200 journal papers, 850 conference papers, and ten books. His research interests include digital image/video analysis and modeling, multimedia data compression, communication and networking, and biological signal/image processing. He is a fellow of The American Association for the Advancement of Science and The International Society for Optical Engineers.